\documentclass[9pt,journal,compsoc]{IEEEtran}

\usepackage{threeparttable}
\usepackage{cite}
\usepackage{amsmath,amssymb,amsfonts}
\usepackage{algorithmic}
\usepackage{graphicx}
\usepackage{textcomp}
\usepackage{indentfirst}
\usepackage{csquotes}
\usepackage{multirow}
\usepackage{tabularx}
\usepackage{float}
\usepackage{color}
\usepackage{comment}
\usepackage[ruled,linesnumbered]{algorithm2e}

\usepackage{multirow}
\newcommand{\squishlist}{
   \begin{list}{$\bullet$}
    { \setlength{\itemsep}{0pt}      \setlength{\parsep}{0pt}
      \setlength{\topsep}{3pt}       \setlength{\partopsep}{0pt}
      \setlength{\listparindent}{-2pt}
      \setlength{\itemindent}{-5pt}
      \setlength{\leftmargin}{1em} \setlength{\labelwidth}{0em}
      \setlength{\labelsep}{0.5em} } }

\newcommand{\squishend}{
    \end{list}  }
    
\usepackage{amssymb}
\usepackage{pifont}

\usepackage[ruled,linesnumbered]{algorithm2e}
\usepackage{pifont}
\usepackage{xcolor}
\usepackage{booktabs}
\usepackage{graphicx}  
\usepackage{stfloats}

\newcommand{\todo}[1]{{\color{red}\sf\bfseries #1}}

\usepackage{comment}
\usepackage{fancyhdr}
\usepackage[normalem]{ulem}
\usepackage{color}
\usepackage{graphicx}
\usepackage{courier}
\usepackage{array}
\renewcommand{\arraystretch}{1.0}
\usepackage{amssymb}
\usepackage{subfig}

\usepackage[normalem]{ulem}

\usepackage[english]{babel}
\usepackage{soul,color,pifont,cite,tikz,balance}

\begin{document}
\title{Characterizing and Understanding GCNs on GPU}
\author{Mingyu Yan\IEEEauthorrefmark{1}\IEEEauthorrefmark{2}\IEEEauthorrefmark{3}, Zhaodong Chen\IEEEauthorrefmark{3}, Lei Deng\IEEEauthorrefmark{3}, Xiaochun Ye\IEEEauthorrefmark{1}, Zhimin Zhang\IEEEauthorrefmark{1}, Dongrui Fan\IEEEauthorrefmark{1}\IEEEauthorrefmark{2} and~Yuan~Xie\IEEEauthorrefmark{3} \\ 
\IEEEauthorrefmark{1}SKLCA, ICT, CAS, Beijing, China,  \IEEEauthorrefmark{2}UCAS, Beijing, China,  \IEEEauthorrefmark{3}UC Santa Barbara, CA, USA
\thanks{
This work was supported in part the National Natural Science Foundation of China 61732018 and NSF 1725447. Corresponding author is Xiaochun Ye.
}
}


\IEEEtitleabstractindextext{%

\begin{abstract}
Graph convolutional neural networks (GCNs) have achieved state-of-the-art performance on graph-structured data analysis. Like traditional neural networks, training and inference of GCNs are accelerated with GPUs. Therefore, characterizing and understanding the execution pattern of GCNs on GPU is important for both software and hardware optimization. Unfortunately, to the best of our knowledge, there is no detailed characterization effort of GCN workloads on GPU. In this paper, we characterize GCN workloads at inference stage and explore GCN models on NVIDIA V100 GPU. Given the characterization and exploration, we propose several useful guidelines for both software optimization and hardware optimization for the efficient execution of GCNs on GPU. 
\end{abstract}

}

\maketitle
\IEEEdisplaynontitleabstractindextext
\IEEEpeerreviewmaketitle

\section{Introduction}

In recent years, Graph Convolutional Neural Networks (GCNs) that operate on graph-structured data have achieved state-of-the-art performance on tasks like node classification, link prediction, and recommendations, etc. GCNs have become a new workload family member in data-centers~\cite{pytorch_biggraph,Graph_Nets_library}. Like traditional neural networks, training and inference of GCN models are accelerated with Graphics Processing Units (GPUs) to achieve an order of magnitude lower latency \cite{pytorch_biggraph}. Therefore, characterizing the execution pattern of GCNs on GPUs is important for both software and hardware optimization for GCNs.

To the best of our knowledge, there is no characterizing effort of GCNs on GPU. Popular GCN models usually contain two major execution phases with distinct execution pattern: \emph{Aggregation} and \emph{Combination}. The former phase aggregates the feature vectors of the neighbor nodes like graph processing, so it exhibits a similar irregular execution pattern. The latter phase updates the feature vectors with multi-layer perceptrons (MLPs), so it has alike regular patterns with traditional neural networks. Nevertheless, GCNs have shown several new features that make their execution patterns differ from traditional workloads, so conclusions in existing characterization studies on graph processing and neural networks cannot be directly inferred in GCNs.

To understand the computation and memory accessing pattern of GCNs, we profile and analyze the inference stage of several GCN models on popular benchmarks with NVIDIA GPU V100, and the results are compared with traditional graph processing and MLP workloads. Besides, we also conduct an exploration of how configurations like dimension size influence the execution time. Our key observations and insights toward architecture design are summarized below.
\squishlist
    \item \textbf{Comparison to Graph Processing:} 1) High-degree spatial data locality and parallelism exist intra vertex; 2) Only inter-warp atomic collision exists; 3) L2 cache hit ratio in \textit{Aggregation} phase is extremely lower than graph processing due to the long reuse distance of vertex data.

    \item \textbf{Comparison to MLP-based Neural Network:} 1) The parameters of MLP exhibit extremely high reusability inter vertex; 2) High-degree parallelism exists inter vertex. 
    
    \item \textbf{Overall Execution:} 1) Hybrid execution pattern exists in GCNs; 2) Execute \textit{Combination} phase ahead of \textit{Aggregation} phase helps reduce data access and computation of \textit{Aggregation} phase; 3) A dataflow exists inter phase for each vertex.
 
    \item \textbf{Exploration:} 1) The execution time of \emph{Combination} is almost proportion to input feature length, while the execution time of \textit{Aggregation} phase in various length are almost the same since it is independent on the length of input feature vector; 2) 
    Both the execution time for \textit{Aggregation} phase and \textit{Combination} phase are almost proportion to output feature length; 3) There are sweet spots for the execution of \textit{Combination} phase in terms of the length of input and output feature.

\squishend

Given the characterization and exploration, we propose useful guidelines as follows for both software framework optimization and hardware optimization for GCNs. 

\squishlist

    \item 
    \textbf{Software Optimization Guideline:} 1) A degree-aware feature access scheduling to reuse the vertex with high degree; 2) Vectorizing atomic operation to improve the efficiency of parallelism; 3) An adaptive execution granularity to leverage the inter-phase dataflow and hardware-optimized function.
 
    \item 
    \textbf{Hardware Optimization Guideline:} 1) A degree- and length-aware replacement policy for Cache to reuse the feature of high-degree vertex and improve memory level parallelism.

\squishend
\section{Background of GCNs}
In general, GCNs follow a neighborhood gather scheme. The feature vector of each vertex is updated by recursive aggregation of the feature vectors of neighbor nodes and combination of features via an MLP or a single fully-connected layer \cite{AliGraph}. Let $h_v^{(k-1)}$ be the feature vector of vertex $v$ at layer $k-1$, $N(v)$ be the neighbor list of vertex $v$, and $\sigma$ be the activation function, we briefly summarize three popular GCN models as follows.

\noindent\textbf{Graph Convolutional Network (GCN) \cite{kipf2016semi}}. The propagation rule of GCN at layer $k$ is defined as follows:
\begin{equation}
    h_v^{(k)}=\sigma\left(mean\left(W^{(k)}h_u ^{(k-1)}|u\in \{N(v)\}\cup \{v\}\right)\right),
\end{equation}
where the term $W^{(k)}h_u ^{(k-1)}$ (\emph{Combination}) multiplies the feature vector of each vertex by the weight matrix $W^{(k)}$ and then the term $mean$ (\emph{Aggregation}) updates each feature vector with the average of its neighborhood.

\noindent\textbf{Graph Isomorphism Network (GIN) \cite{xu2018powerful}}. In GIN-0 introduced in Xu et al. (2018) \cite{xu2018powerful}, the feature vector of each vertex is updated with
\begin{equation}
    h_v^{(k)}\!=\!mlp\left(sum\left(h_u ^{(k-1)}|u\in \{N(v)\}\cup \{v\}\right)\right),
\end{equation}
in which the feature vectors are first aggregated by the summation of neighborhood and then updated with MLP.

\noindent\textbf{GraphSAGE (SAG)\cite{hamilton2017inductive}}. In GraphSAGE, the feature vectors are updated with the same propagation rule of \textbf{GCN}, the difference is that while \textbf{GCN} updates the feature vectors of all vertexes in the graph in each iteration, GraphSAGE only update a batch of vertexes along with their 2-hop neighbors in an iteration.

There are three major different features between GCNs with traditional graph processing and neural networks:
\begin{enumerate}
    \item \textit{\uline{Large and Variable Feature Length}}. The feature data in graph processing are small, usually one element for each vertex, while the feature vector of each vertex in the \emph{Aggregation} phase of GCNs usually contains hundreds of entries and varies across layers and datasets.
    \item \textit{\uline{Parameters Shared by Vertices}}. In traditional MLP-based neural network, to classify one sample, only one feature vector is forward through the MLP, and the parameters in the MLP are not shared. 
    However, in node classification of GCNs with $k$ layers, the feature vectors of all $k$-hop neighbours are forwarded; In graph classification, the feature vectors of all vertexes are required. 
    As a result, in GCNs, the parameters in the MLP can be fully shared by each feature vector. 
    \item \textit{\uline{Alternative Execution}}, the two phases are executed alternatively until the final result is produced. 
\end{enumerate}

\section{Evaluation Setup}

\noindent\textbf{Benchmark.} Table \ref{table:data_set} and Table \ref{table:gcn_model} provide the information of the benchmark GCN models and  graph datasets used in our evaluation. For GCNs, we select three advanced models: Graph Convolutional Network (GCN) \cite{kipf2016semi}, Graph Isomorphism Network (GIN) \cite{xu2018powerful}, and GraphSAGE (SAG) \cite{hamilton2017inductive}. For clarity, we evaluate the first graph-convolutional layer of each model on popular datasets including Cora, Citeseer, Pubmed, and Reddit. For classical graph processing, we run PageRank on the Reddit and LiveJournal dataset. For traditional MLP, we test a single fully-connected layer on MNIST. Notably, we mainly focus on the inference stage rather than training.

\begin{table}[!htbp]
	\caption{Configuration of convolution layers. Here $|h_u^{(k-1)}|$ denotes the length of feature vector $h_u^{(k-1)}$.}\label{table:gcn_model}
	\vspace{-5pt}
	\centering
\renewcommand\arraystretch{1.3}
  \resizebox{0.48\textwidth}{!}{
\begin{tabular}{ccc}
\toprule
                                        &  \textbf{Aggregation Operator \& Combination Operator}                              \\ \midrule
\textbf{GCN (GCN)}                      &  Mean  \& MLP: $|h_u^{(k-1)}|$--128                 \\ 
\textbf{GraphSage (SAG)}                &  Mean \& MLP: $|h_u^{(k-1)}|$--128                                   \\ 
\textbf{GINConv (GIN)}                  &  Add  \& MLP: $|h_u^{(k-1)}|$--128--128                              \\ \midrule

\textbf{PageRank (PGR)}                 &  Graph Processing                                      \\ 
\textbf{MLP-MNIST}                      &  MLP: 784--128 with batching size 1000                              \\ \midrule

\end{tabular}
}
\end{table}

\vspace{-10pt}
\begin{table}[!htpb]
	\caption{Datasets information \cite{1stChebNet,GraphDynS}}\label{table:data_set}
	\centering
	\vspace{-5pt}
\renewcommand\arraystretch{1.3}
  \resizebox{0.40\textwidth}{!}{
		\begin{tabular}{*5{c}}
	     	\toprule   
			\textbf{Dataset}           & \textbf{\#Vertex}  &  \textbf{Feature Len.}        & \textbf{\#Edge}      \\   
			 \textbf{Cora (CR)}        & 2,708       & 1,433           & 5,429           \\  
			 \textbf{Citeseer (CS)}    & 3,327       & 3,703     	   & 4,732           \\  
  			 \textbf{Pubmed (PB)}      & 19,717      & 500             & 44,338          \\ 
  			 \textbf{Reddit (RD)}      & 232,965     & 602             & 11,606,919       \\
             \textbf{LiveJournal (LJ)} & 4,847,571   & 1               & 68,993,773       \\

             \bottomrule
		\end{tabular}
}	
\end{table}

\noindent\textbf{Profiling Platform.} The GCN models are implemented with the state-of-the-art GPU-based software framework for GCNs: \emph{PyTorch Geometric} \cite{PyTorch_Geometric}. The PageRank is implemented with Gunrock \cite{Gunrock}. All the workloads are profiled on single NVIDIA GPU V100 with NVIDIA NVProf and averaged among 5 iteration.
\section{Observation and Analysis}

This section is organized as follows. First, we present an overview of our profiling result. Then, we characterize dominant kernels in \emph{Aggregation} and \emph{Combination} phase and compare them with traditional workloads. At last, we explore the impact of feature length on execution time.

\subsection{Overview of Profile}
\begin{figure}[!hptb] 
    \centering
    \includegraphics[page=1, width=\linewidth]{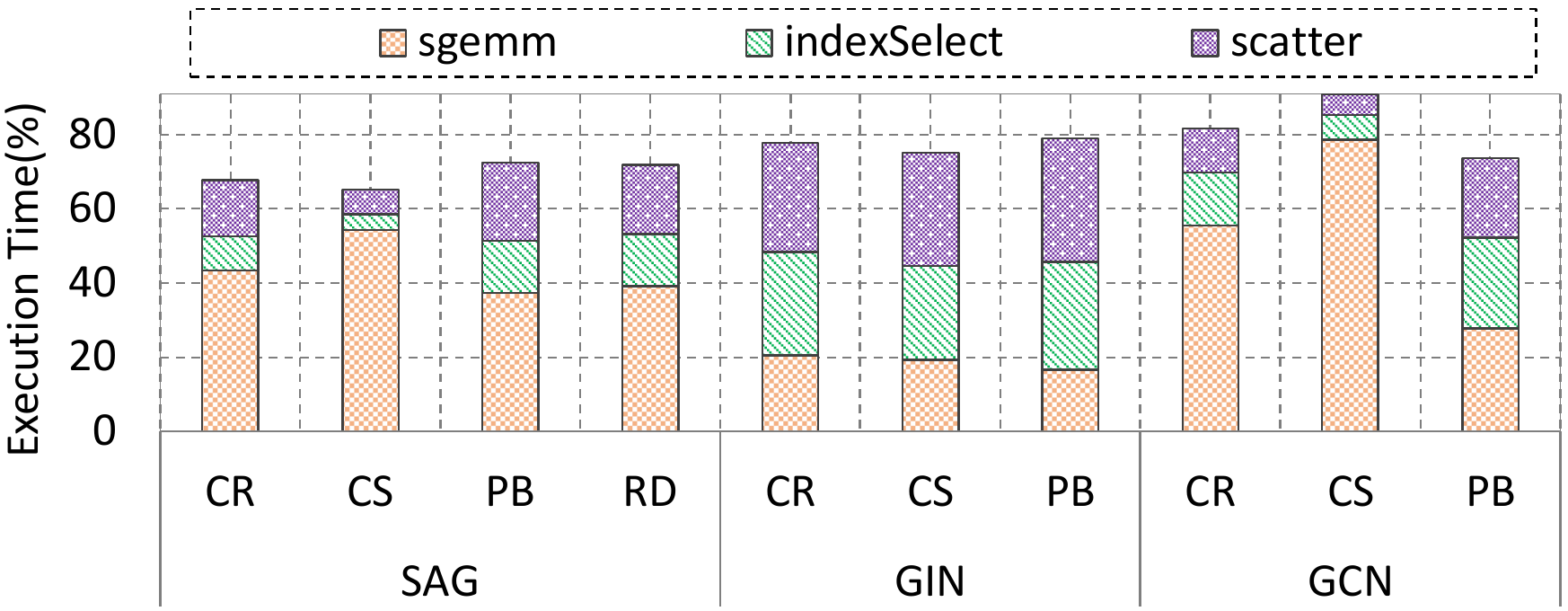}
    \caption{Execution time breakdown on V100 GPU.}
    \vspace{-20pt}
    \label{fig:execution_time_breakdown}
\end{figure}

\textbf{Execution Time Breakdown.} 
Fig.~\ref{fig:execution_time_breakdown} illustrates the execution time breakdown of major kernels that occupy most of execution time on GPU.

The \emph{sgemm} kernels multiplies the weight matrix with feature vectors to perform \emph{Combination}. The \textit{indexSelect} kernel and \textit{scatter} kernel execute \textit{Aggregation} function for all vertices.
Specifically, the \textit{indexSelect} kernel uses the neighbor ID to select the neighbor's feature vector of each vertex, and then uses these feature vectors to build a dense feature matrix for the input of \textit{scatter} kernel.
Each thread in the \textit{scatter} kernel executes aggregation operator for each element in a neighbor's feature vector. 

As illustrated in Fig. \ref{fig:execution_time_breakdown}, the above three kernels take up 65\% to 90\% execution time in different configurations. The portion of execution time that each kernel takes is determined by the sequence of \emph{Combination} and \emph{Aggregation} as well as the length of feature vectors. Specifically, GIN executes \emph{Aggregation} phase first while the other two execute \emph{Combination} phase first. While the scale of \emph{Combination} is similar among the three models, the feature vector length in GCN and SAG are significantly reduce by \emph{Combination}, so their \emph{Aggregation} phase take much fewer time than GIN. In terms of dataset, 
the \emph{Combination} takes more execution time in the datasets with longer feature length, i.e. CS.
\vspace{-10pt}

\subsection{Analysis of \textit{Aggregation} Phase} \label{sec:analysis_of_aggr}

Here, we provide detailed analysis of the \emph{Aggregation} phase in SAG and compare it with PGR on the RD and LJ datasets.

\begin{figure*}[!htbp] 
    \centering
    \begin{minipage}{0.82\textwidth}
    \centering
    \includegraphics[page=1, width=\textwidth]{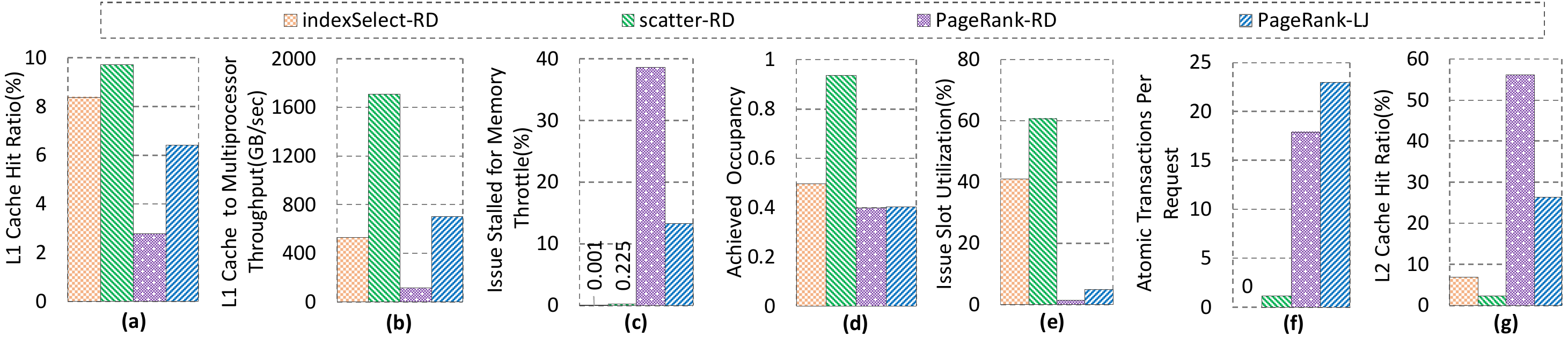}
    \vspace{-20pt}
    \caption{The profiling result of \textit{Aggregation} phase.}
    \label{fig:aggregation_analysis}
    \end{minipage}\hfill
    \begin{minipage}{0.18\textwidth}
    \centering
    \includegraphics[page=1, width=\textwidth]{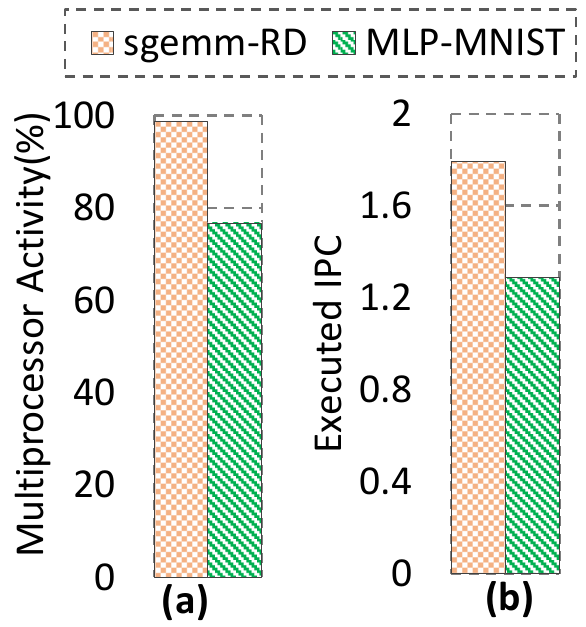}
    \vspace{-20pt}
    \caption{The result of \textit{Combination} phase.}
    \label{fig:combination_analysis}
    \end{minipage}\hfill
    \vspace{-20pt}
\end{figure*}

\uline{High-degree spatial data locality exists in the access to feature data.} This locality derives from the access to each long-length feature vector. 
Fig.~\ref{fig:aggregation_analysis} (a) and (b) respectively illustrate that \textbf{L1 cache hit ratio} and \textbf{L1 cache to multiprocessor throughput(GB/sec)} of \textit{Aggregation} phase are higher than that in graph processing on the same dataset (i.e., RD). Besides, Fig.~\ref{fig:aggregation_analysis}(c) depicts that the value of \textbf{Memory Throttle} of PGR is extremely higher than \textit{Aggregation} phase (39.27\% v.s. 0.225\%). It means a large number of pending memory operations prevent further forward progress in the micro-architecture pipeline on PGR due to the fine-grained and irregular data access to feature data. However, the accesses to long-length feature data in \textit{Aggregation} phase can be reduced by combining several memory transactions into one. 
These three results demonstrate that \textit{Aggregation} phase has an extra spatial locality in the access to feature data compared to graph processing.

\uline{High-degree parallelism exists intra vertex.} Except for the inter-vertex and inter-edge parallelism as in graph processing, \textit{Aggregation} phase possesses a new kind of parallelism, the intra-vertex parallelism. This parallelism comes from the element-wise aggregation of each neighboring vertex feature vector. As a result, the \textbf{Achieved Occupancy} and \textbf{Issue Slot Utilization} of \textit{Aggregation} phase are higher than that in graph processing as shown in Fig.~\ref{fig:aggregation_analysis}(d) and (e). 

\uline{Only inter-warp atomic collision exists in parallelism exploitation.}
Atomic collision refers to scenarios where multiple threads try to read-modify-write the same data word simultaneously. To guarantee the atomicity, these updates from different threads will be serialized.
There are two collisions existing in GPUs: inter-warp collision and intra-warp collision. 
Since each thread inside a warp processes one of the consecutive feature elements of neighboring feature vector, there is almost no intra-warp collision in \textit{Aggregation} phase. 
In contrary, in PGR, each thread processes a random vertex with a single feature element.
It means that the threads intra or inter warp may update the same vertex, which causes inter-warp collision or intra-warp collision. 
Thus, \textbf{Atomic Transactions Per Request} is 1.1 in \textit{Aggregation} phase (Fig.~\ref{fig:aggregation_analysis}(f)), smaller than the 17.9 in PGR. 

\uline{\textit{Aggregation} phase exhibits lower reuse of neighbors' feature data than that in graph processing.}
Although the graph traversal in \textit{Aggregation} phase is same to that in PGR, \textbf{L2 cache hit ratio} in \textit{Aggregation} phase is extremely low.
Fig.~\ref{fig:aggregation_analysis}(g) shows \textbf{L2 cache hit ratio} of \textit{Aggregation} phase is only 6.9\% while that is 56.2\% in PGR, even although they process the same graph. The reason is as follows.
The vertex data in graph processing is only one element, while the feature vector contains hundreds element in \textit{Aggregation} phase. 
As a result, L2 cache can hold many vertex data in PGR, which enlarges the opportunity to reuse the vertex data of the shared neighbor. 
However, L2 cache can only contain smaller amount of feature vectors in \textit{Aggregation} phase, which results in longer data reuse distance of feature vector than that in graph processing. Therefore, \textit{Aggregation} phase exhibits low reuse of neighbors' feature data.

\subsection{Analysis of \textit{Combination} Phase}

Here, we provide detailed analysis of the \emph{Combination} phase on SAG model with RD dataset and compare it with MLP-MNIST with batch size 1,000, such that both model has similar batch size and input feature vector length.

\uline{The parameters of neural network exhibits extremely high reusability inter vertex.} To classify one handwritten number in MNIST with MLP, only a single feature vector need to be forwarded. On the contrary, in node classification tasks, the feature vectors of all the neighbors within $k$-hop of the target node need to be processed. In graph classification tasks, all the vertex in the graphs should be processed. When processing multiple feature vectors, the parameters in the model can be shared across all the features. As a result, \textit{Combination} phase presents more reusability on the parameters of neural network. 

\uline{High-degree parallelism exists inter vertex.} As mentioned before, the amount of feature vectors processed in \emph{Combination} phase is much larger than traditional neural networks, which introduces more parallelism. As a result, it is more capable of feeding up the thousands of parallel floating-point units in GPU and hiding the latency to memory. As shown in Fig.~\ref{fig:combination_analysis} (a) and (b), \textbf{Multiprocessor Activity(\%)} and \textbf{Executed IPC} of \textit{Combination} phase are 98.885\% and 1.8, more than 76.829\% and 1.3 respectively, the values of MLP-MNIST.

\subsection{Analysis of Overall Execution}
Here, we analysis the overall execution on SAG model.
\vspace{-10pt}
\begin{figure}[!hptb] 
    \centering
    \includegraphics[page=1, width=\linewidth]{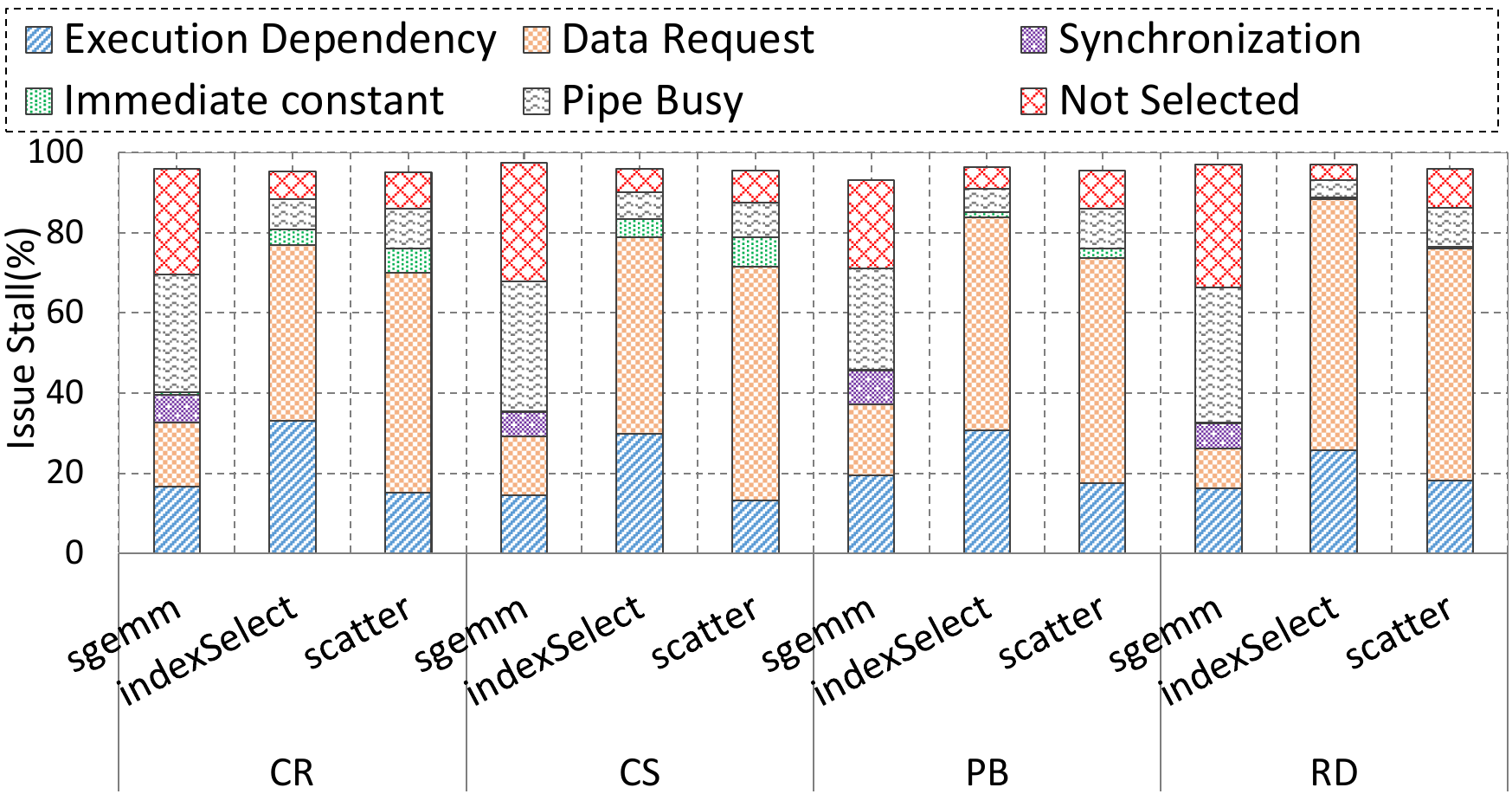}
    \caption{Percentages of issue stall reasons on SAG model.}
    \label{fig:stall_reasons}
\end{figure}
\vspace{-5pt}

\uline{Hybrid execution patterns exist.} In \emph{Aggregation} phase, the \textbf{Computation Unit Utilization} is only 50\% and the \textbf{Executed IPC} is only 1.78 on average as shown in Table \ref{tab:characterization}. The aggregation heavily relies on the graph structure so that it is obstructed by irregularity~\cite{GraphDynS} and load-load data dependency chain \cite{Memory_Hierarchy_Graph}. Therefore, it is mainly stalled for \textit{Data Request} and \textit{Execution Dependency} as depicted in Fig.~\ref{fig:stall_reasons}. The irregularity also leads to low \textbf{L2 Cache Hit Rate} (6.87\%) and high \textbf{DRAM Byte per Operation} (2.35). In contrary, the \emph{Combination} phase achieves 90\% \textbf{Computation Unit Utilization} and 2.49 \textbf{Executed IPC}. The intensive Float-Point calculations well hide the data access latency, and the stalls are issued majorly for \textit{Pipe Busy} and \textit{Not Selected}, which is due to the limited number of computation units. The regular execution pattern leads to high spatial and temporal data locality, the \textbf{L2 Cache Hit Rate} is 82.5\% and \textbf{DRAM Byte per Operation} is as low as 0.01. As a result, while the \emph{Aggregation} phase is memory bound with irregular data access pattern and low data reusability, the \emph{Combination} phase is computation bound with regular data access pattern and high data reusability.

\begin{table}[!hptb]
\vspace{-5pt}
\scriptsize
\centering
\caption{Characterization of hybrid execution patterns on RD}\label{tab:characterization}
      \renewcommand\arraystretch{1.2}
      \resizebox{0.42\textwidth}{!}{
      \begin{tabular}{ccc}
      \toprule
                                             & \textbf{\emph{Aggregation}}    &  \textbf{\emph{Combination}}    \\ \midrule
      \textbf{Computation Unit Utilization}  & 50\%                           & 90\%\\
      \textbf{Executed IPC} & 1.78 & 2.49\\
      \textbf{L2 Cache Hit Rate}             & 6.87\%                          &   82.5\%  \\   
      \textbf{DRAM Byte per Operation}       & 2.35                            &   0.01 \\ 
      \midrule
      \textbf{Execution Bound}               & Memory                         &   Compute \\ 
      \textbf{Data Access Pattern}           & Irregular                 &   Regular      \\         
      \textbf{Data Reusability}              & Low                            &   High              \\ 
      \bottomrule
      \end{tabular}}
      \vspace{-5pt}
\end{table}

\uline{Execute \textit{Combination} phase ahead of \textit{Aggregation} phase helps reduce data access and computation of \textit{Aggregation} phase.} While the feature length in RD is 602, the \emph{Combination} phase usually reduces the dimension to 128 by a factor of $4.7\times$. Therefore, in the \textit{Aggregation} phase, the data access to neighbor's feature vector becomes less and the computation for the aggregation of each neighbor also becomes less.  
Table.~\ref{tab:reduction_access_computation} illustrates the reduction of data accesses and computations in \textit{Aggregation} phase, up to 4.75$\times$ and 4.72$\times$ respectively. Moreover, the performance achieves 4.76$\times$ improvement.
\vspace{-7pt}
\begin{table}[!hptb]
      \scriptsize
      \centering
      \caption{Impact of the execution flow on \textit{Aggregation} phase.}\label{tab:reduction_access_computation}
      \vspace{1pt}
      \renewcommand\arraystretch{1.2}
      \resizebox{0.48\textwidth}{!}{
      \begin{tabular}{ccccccc}
      \toprule
                  & \multicolumn{2}{c}{\textit{\textbf{Com $\rightarrow$ Agg}}} & \multicolumn{2}{c}{\textit{\textbf{Agg $\rightarrow$ Com}}} & \multicolumn{2}{c}{\textit{\textbf{Reduction}}} \\  \midrule
    \textbf{Data Accesses (bytes)}      &   \multicolumn{2}{c}{568,064,375}  & \multicolumn{2}{c}{2,698,865,170}     & \multicolumn{2}{c}{4.75$\times$}
    \\
    \textbf{Computations (Operations)}  &   \multicolumn{2}{c}{231,995,186}  &   \multicolumn{2}{c}{1,096,220,688}  & \multicolumn{2}{c}{4.72$\times$}
    \\       
    \textbf{Execution Time (ms)}  &   \multicolumn{2}{c}{1.12}  &   \multicolumn{2}{c}{5.34}  & \multicolumn{2}{c}{4.76$\times$} \\   
    \bottomrule
      \end{tabular}}
     \vspace{-5pt}
\end{table}

\uline{A dataflow exists inter phase in GCNs for each vertex.} 
The result of each vertex in \textit{Aggregation} phase is taken as the input of \textit{Combination} phase for the transformation of each vertex. It indicates that a vertex is able to start the execution in \textit{Combination} phase after this vertex completes its aggregation.
Therefore, an inter-phase dataflow exists in GCNs for each vertex. 
However, to leverage the hardware-optimized functions, the implementation of GCNs on GPU misses this inter-phase dataflow. As a result, many unnecessary data accesses and data addressing computations are introduced.


\subsection{Exploring GCN Model}

Here, we explore the new features of GCNs on SAG model with RD dataset. As SAG executes \emph{Combination} phase ahead of \emph{Aggregation} phase, the execution time of \emph{Combination} phase is determined by the length of both input and output feature vector, while \emph{Aggregation} is only determined by the length of output feature vector.
\begin{figure}[!hptb] 
    \vspace{-8pt}
    \centering
    \includegraphics[page=1, width=0.99\linewidth]{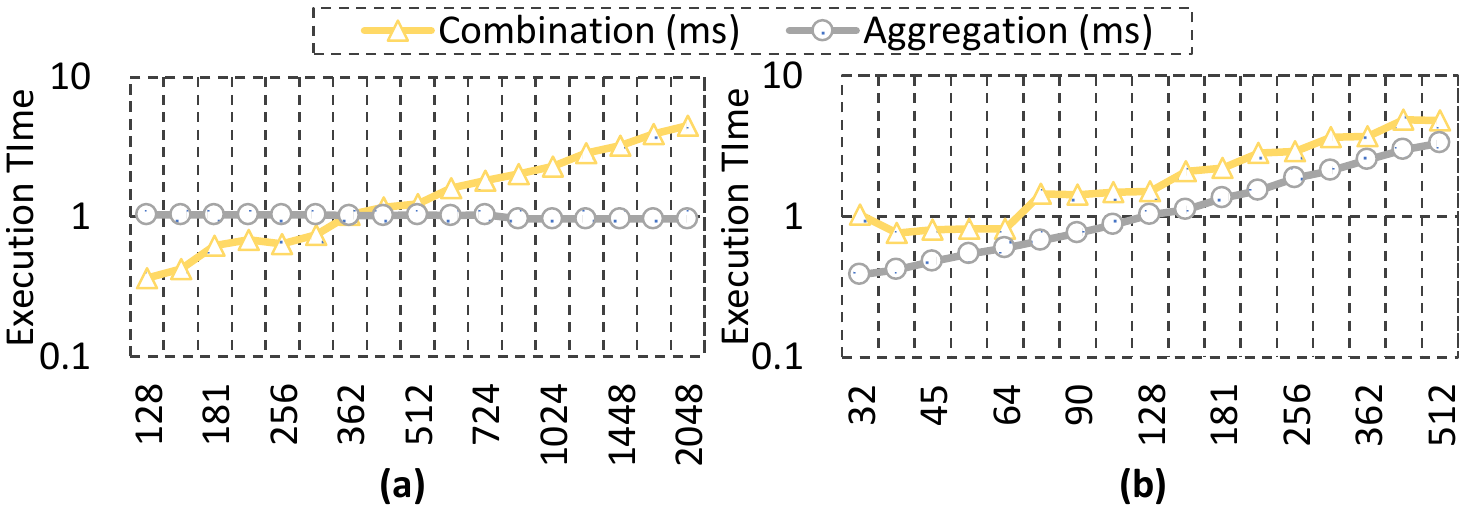}
    \vspace{-5pt}
    \caption{Exploration on the length of input feature vector (a) and output feature vector (b).}
    \label{fig:exploartion}
    \vspace{-5pt}
\end{figure}

\noindent\textbf{Various Length of Input Feature Vector.} As illustrated in Fig. \ref{fig:exploartion} (a), the execution time of \emph{Combination} phase is almost proportion to input feature length. An interesting observation is that there are sweet spots when the input dimension is the index of 2, i.e. 256.

\noindent\textbf{Various Length of Output Feature Vector.} As illustrated in Fig. \ref{fig:exploartion} (b), the execution time of \emph{Aggregation} phase increases linearly with the output feature length. On the other hand, the \textit{Combination} phase is insensitive to the output feature size when the feature length is smaller than 64, which is due to the redundant computational resources. Besides, the sweet spot still exists when the output dimension is the index of 2.

\section{Architectural Guidelines} 

\subsection{Software Optimization Guideline}

\textbf{Degree-aware Feature Access Scheduling.} Real-world graphs possess well-connected regions where relatively few vertices share edges with many common neighbors. It indicates that the vertices with large degree exhibits high reusability on their feature data. Thus, an online data access scheduling can leverage that to shorten the reuse distance. 

\textbf{Vectorizing Atomic Operation.} To improve the parallelism efficiency, vectorizing atomic operation is available for \textit{Aggregation} to reduces the atomic overhead in GPU since only inter-warp collision exits in GCNs.

\textbf{Adaptive Execution Granularity.} Leveraging inter-phase dataflow is an excellent opportunity to overlap the memory-bound \textit{Aggregation} phase and computation-bound \textit{Combination} phase. Meanwhile, it is also important to leverage the architecture's advantage. Therefore, an appropriate or adaptive granularity for execution can achieve a better trade-off. 

\subsection{Hardware Optimization Guideline}

\textbf{Degree- and Length-aware Replacement Policy.} To ease the programmer efforts and improve data reuse, L2 Cache can be modified to equip a degree- and length-aware replacement policy. This policy can replace the vertex feature by aware of its degree, which indicates its reusability. Besides, it can replace the whole vertex feature vector in a time since all the elements in vector are used together. This way helps fire many requests at the same time to exploit the high bandwidth memory.


\section{Conclusion} 
In this work, we characterize and explore an emerging application GCNs on NVIDIA V100 GPU. The characterization results can help programmers understand the execution pattern of GCNs. We also believe the observations made in this paper will provide useful guidance to enable future architecture and system research for GCNs.

\vspace{0pt}{
\scriptsize

\bibliographystyle{ieeetr}
\bibliography{ref.bib}}
\end{document}